\documentclass{llncs}

\usepackage[english]{babel}
\usepackage[utf8]{inputenc}
\usepackage{amsmath}
\usepackage{graphicx}
\graphicspath{{imgs/}{../imgs/}}
\usepackage{epstopdf}
\usepackage{subcaption}
\usepackage{float}
\usepackage{mathtools}

\usepackage[hidelinks]{hyperref}
\hypersetup{
  colorlinks = true,
  urlcolor   = blue,
  linkcolor  = blue,
  citecolor  = red
}

\title{Cyclic Dominance in the Spatial Coevolutionary Optional Prisoner's Dilemma Game
\thanks{The final publication will be available in the Proceedings of the 24th Irish Conference
on Artificial Intelligence and Cognitive Science (AICS 2016), Dublin, Ireland, September 20-21, 2016.
\url{http://ceur-ws.org/Vol-1751}}
}

\author{
    Marcos Cardinot \and
    Josephine Griffith \and
    Colm O'Riordan
}
\institute{
    Information Technology, National University of Ireland, Galway, Ireland \\
    \email{marcos.cardinot@nuigalway.ie}
}

\begin{document}
\maketitle

\begin{abstract}
This paper studies scenarios of cyclic dominance in a coevolutionary spatial
model in which game strategies and links between agents adaptively evolve over
time. The Optional Prisoner's Dilemma (OPD) game is employed. The OPD is an
extended version of the traditional Prisoner's Dilemma where players have a
third option to abstain from playing the game. We adopt an agent-based
simulation approach and use Monte Carlo methods to perform the OPD with
coevolutionary rules. The necessary conditions to break the scenarios of cyclic
dominance are also investigated. This work highlights that cyclic dominance is
essential in the sustenance of biodiversity. Moreover, we also discuss the
importance of a spatial coevolutionary model in maintaining cyclic dominance
in adverse conditions.
\end{abstract}

\section{Introduction}

Competition is one of the most fundamental concepts in the study of the
interaction between individuals in an ecosystem.  Competition occurs when there
is a contest for resources, such as food, mates or territories. Competition
also favours a selection process, which reaches its peak with the dominance of
better-adapted individuals and the extinction of less-adapted individuals. In
many cases, rather than one individual dominating, the system can reach an
equilibrium where individuals can coexist \cite{Jorgensen2014}.

Scenarios of coexistence may occur when two or more individuals, or species,
form a cycle of dominance. For instance, considering a population of three
species: $X$, $Y$ and $Z$; cyclic dominance occurs when $X$ dominates $Y$,
$Y$ dominates $Z$, and $Z$ dominates $X$, forming a closed loop, which is
also known as an intransitivity \cite{Szolnoki2014}.

In nature, cyclic dominance plays a key role in the sustenance of biodiversity.
For example, the male side-blotched lizard shows an intransitive behaviour when
guarding their mates. This kind of lizard can be divided into three categories
based on their throat colours:
\begin{itemize}
    \item Blue-throated males guard small territories with a single female.
        They are efficient in defending their mate from yellow-throated lizards.
    \item Yellow-throated males do not guard territories at all, but they move
        around in search of mates.
    \item Orange-throated males guard larger territories, keeping harems of
        females. Consequently, as they have to split their efforts defending
        several territories, they are less efficient in defending them from
        yellow-throated lizards. However, they are more aggressive and can
        steal mates from blue-throated lizards.
\end{itemize}
Therefore, a cyclic competition exists because the blue-throated males beat
the yellow-throated males, the yellow-throated males beat the orange-throated
males, and the orange-throated males beat the blue-throated males
\cite{Sinervo2000}.

It is noteworthy that as the number of species, or the population size,
increases, the collective behaviours of a system subject to these sort of
scenarios may become much harder to analyse and predict.  In this way,
frameworks like evolutionary game theory, which employs game theory to evolve
populations of rational agents, have been widely applied by researchers as it
provides many useful insights to explain such a complex scenario
\cite{Smith1982}.

Particularly, games such as the Rock-Paper-Scissors game
\cite{Yu2016,Mobilia2010,Cheng2014,Reichenbach2007,Kerr2002} and the Prisoner's
Dilemma game \cite{Yu2015,Jeong2014,Szolnoki2010,Szabo2002} have been studied
in the context of cyclic competition.
In these games, a participant's interactions are generally constrained by
particular graph topologies \cite{Nowak1992}, where it has been shown that the
spatial organisation of strategies may also affect the
outcomes~\cite{Cardinot2016sab}.
Recent studies have also explored dynamically weighted networks,
where it has been shown that the coevolution of both the game strategies and the
spatial environment can further help in understanding real-world systems
\cite{Cardinot2016ecta,Huang2015,Wang2014,Demirel2011,Zimmermann2001}.

In this paper, we employ extensive agent-based Monte Carlo simulations to
perform the coevolutionary Optional Prisoner's Dilemma game in a population of
agents placed on a lattice grid where, game strategies, and the edges linking
agents, adaptively evolve over time. We aim to investigate:
\begin{itemize}
    \item scenarios of cyclic dominance in the Optional Prisoner's Dilemma, and
    \item the necessary conditions needed to break the scenarios of cyclic dominance.
\end{itemize}
Specifically, the experiments performed are:
\begin{itemize}
    \item Finding scenarios of cyclic dominance.
    \item Exploring the sustenance of coexistence after the extinction of one strategy.
    \item Investigating the robustness of the coexistence of three strategies when
        some portion of the strategies mutate into another strategy.
    \item Investigating the impact of the coevolutionary rules in the
        sustenance of coexistence.
\end{itemize}

The paper outline is as follows: Section~\ref{sec:related} presents a brief
overview of the previous work in cyclic dominance in the context of
evolutionary and coevolutionary game theory.  Section~\ref{sec:methodology}
introduces the Optional Prisoner's Dilemma game, describes the coevolutionary
model adopted, and outlines the experimental set-up.  In
Section~\ref{sec:cyclic}, some scenarios of cyclic competition between three
strategies are investigated. Section~\ref{sec:extinction} explores the necessary
conditions to break the cyclic competition.  Lastly,
Section~\ref{sec:conclusion} summarizes the main results and outlines future
work.

\section{Related Work}
\label{sec:related}

Started by John Maynard Smith, evolutionary game theory has been studied since
the 1980s where ideas from evolutionary theory have been applied to game
theory. Game theory models situations of conflict between rational agents,
i.e., individual players make decisions, in which the outcome will depend on
the other players' decisions \cite{Smith1982}.

Evolutionary game theory has been used as an important framework to explore and
study the phenomena of cyclic competition, or intransitivity, which can be
found in many real-world systems in different domains such as biology
\cite{Reichenbach2007,Kerr2002} and physics \cite{Knebel2015,Cheng2014}.
Moreover, it has been widely studied at a higher level of abstraction,
providing insights into the understanding of oscillatory and stochastic systems
\cite{Yu2016,Mobilia2010}.

Despite the fact that cyclic competition has been observed in two-strategy games
\cite{Szolnoki2010}, it is more likely to happen in games involving three or
more strategies \cite{Szolnoki2014}. The rock-paper-scissors (RPS) game
remains one of the most oft-studied games in scenarios of cyclic dominance due
to its intransitive nature, in which the loop of preference between pairs of
strategies is very straightforward --- paper covers rock, rock crushes scissors
and scissors cuts paper \cite{Zhou2016,Juul2013}.
It is noteworthy that such an intransitive behaviour has also been noticed in
other evolutionary games such as the Optional Prisoner's Dilemma game
\cite{Jeong2014,Szabo2002} and the voluntary public goods game
\cite{Hauert2008}. However, they have been much less explored in this specific
context of cyclic dominance.

For instance, Yu et al. \cite{Yu2015} proposed a study of the influence of the
population size and the level of individual rationality on the evolutionary
dynamics of the Voluntary Prisoner's Dilemma (VPD) game, which is very similar
to the Optional Prisoner's Dilemma (OPD) game, in which a third type of
strategy is considered. In their paper, scenarios of cyclic dominance in the
VPD game are discussed. It was shown that these scenarios prevent  the full
dominance of a specific strategy in the population.

Recent studies have also explored the use of coevolutionary rules in game
theory. First introduced by Zimmermann et al. \cite{Zimmermann2001}, those
rules  propose a new model in which agents can adapt their
neighbourhood during a dynamical evolution of graph topology and game strategy.
As discussed by Perc and Szolnoki \cite{Perc2010}, the coevolutionary games
constitute a natural upgrade of the well-know spatial evolutionary games
\cite{Nowak1992}, where dynamic spatial environments are taken into
consideration. In fact, the coevolution of strategies and spatial environment
has given ground to a new trend in evolutionary game theory due to its wider
applicability in the understanding of more realistic scenarios
\cite{Huang2015,Wang2014}.

The inclusion of coevolutionary rules in the Optional Prisoner's Dilemma game
has been recently proposed by Cardinot et al. \cite{Cardinot2016ecta}, who
identified that coevolutionary rules may favour the emergence of cyclic
competition. However, many questions remain unanswered, such as its robustness
against frozen states, i.e., when strategies become extinct because of some
disturbance in the system.

\section{Methodology}
\label{sec:methodology}

In this section, we will describe the Optional Prisoner's Dilemma game, the
Monte Carlo methods and the coevolutionary rules adopted. Finally, the spatial
environment and the experimental set-up are outlined.

\subsection{The Optional Prisoner's Dilemma}

The Optional Prisoner's Dilemma (OPD) game is an extension of the classical
version of the Prisoner's Dilemma (PD) game. This extension incorporates the
concept of abstinence, where agents can abstain from playing the game. It leads
to a three-strategy game in which agents can not only defect or cooperate, as
in the classical PD, but can also choose to abstain from a game interaction.
Consequently, there are nine payoffs associated with each pairwise interaction
between strategies. However, as defined in other studies
\cite{Cardinot2016sab,Hauert2008}, in this work we also assume that if one or
both players abstain, both will obtain the same payoff, which is called the
loner's payoff ($L$). Hence, as illustrated in Table~\ref{tab:payoff1}, the
OPD game is actually characterised by five payoffs, where the other four
payoffs are known as the reward for mutual cooperation ($R$), punishment for
mutual defection ($P$), sucker's payoff ($S$) and temptation to defect ($T$).

\begin{table}[htb]
    \vspace{-20px}
    \centering
    \caption{The Optional Prisoner's Dilemma game matrix.}
    \label{tab:payoff1}
    \setlength{\tabcolsep}{10pt}
    \begin{tabular}{c| c c c}
            & {\bf C} & {\bf D} & {\bf A} \\
        \hline
        {\bf C} & R,R     & S,T     & L       \\
        {\bf D} & T,S     & P,P     & L       \\
        {\bf A} & L       & L       & L       \\
    \end{tabular}
    \vspace{-10px}
\end{table}

In order to establish the dilemma of the OPD, it is important to
consider that the loner's payoff ($L$) obtained by abstainers is greater than
$P$ and less than $R$, and that the traditional constraints of the Prisoner's
Dilemma still hold, i.e., $T>R>P>S$. Thus, in this extension the dilemma arises
when the payoff values are ordered such that $T>R>L>P>S$. In consonance with
common practice \cite{Yu2015,Huang2015,Nowak1992}, as the evolutionary rule
depends on the payoff differences between agents, the payoff values can be
rescaled to $R=1$, $P=0$, $S=0$, $T=b$ and $L=l$, where $1<b<2$ and $0<l<1$,
which, in turn, maintain the dilemma.

\subsection{Monte Carlo Simulation}

This work considers a population of $N$ agents placed on a square lattice with
periodic boundary conditions, i.e., a torus topology (upper-bottom and
left-right borders must match each other exactly). In this lattice, each agent
interacts only with its eight immediate neighbours (Moore neighbourhood) by
playing the Optional Prisoner's Dilemma game with coevolutionary rules. In our
experiments, initially, each agent is designated as an abstainer ($A$),
cooperator ($C$) or defector ($D$) with equal probability. Each edge linking
agents has the same weight $w=1$, which will adaptively change in accordance
with the agents' interactions.

Monte Carlo simulations are performed to investigate the dynamics of the
coevolution of both game strategy and link weights. In one Monte Carlo (MC)
step, each player is selected once on average, that is, one MC step comprises
$N$ inner steps where the following calculations and updates occur: an agent
($x$) is randomly selected from the population; its utility $u_{xy} = w_{xy} P_{xy}$
is calculated for each of its eight neighbours (represented as $y$), where
$w_{xy}$ is the link weight between agents $x$ and $y$, and $P_{xy}$
corresponds to the payoff obtained by agent $x$ on playing the game with agent $y$;
the average accumulated utility, i.e. ${\bar{U_x}=\sum u_{xy} / 8}$, is
calculated and used to update the link weights (Eq.~\ref{eq:link}); as the link
weights have been updated, the utilities
must be recalculated; finally, strategies are updated based on the
comparison of the accumulated utilities $U_x$ and $U_y$ (obtained from a
randomly selected neighbour) (Eq.~\ref{eq:prob}).


As shown in Equation \ref{eq:link}, the link weight ($w_{xy}$) between agents is
updated by comparing the utility ($u_{xy}$) and the average accumulated utility
($\bar{U_x}$),
\begin{equation}\label{eq:link}
    w_{xy} =
    \begin{dcases*}
        w_{xy} + \Delta  & if $u_{xy} > \bar{U_x}$ \\
        w_{xy} - \Delta  & if $u_{xy} < \bar{U_x}$ \\
        w_{xy}           & otherwise
    \end{dcases*},
\end{equation}
where $\Delta$ is a constant such that ${0 \le \Delta \le \delta}$, where
$\delta$ (${0 < \delta \le 1}$) defines the weight heterogeneity.
Moreover, as done in previous research \cite{Cardinot2016ecta,Huang2015,Wang2014},
the link weight $w_{xy}$ is also adjusted to be within the range of
${1-\delta}$ to ${1+\delta}$. In this way, when $\Delta=0$ or $\delta=0$, the
link weight keeps constant (${w=1}$), which results in the traditional scenario
where only the strategies evolve.

In order to update the strategy of the agent $x$, the accumulated utilities
$U_x$ and $U_y$ are compared such that, if $U_y>U_x$, agent $x$ will copy the
strategy of agent $y$ with a probability proportional to the utility difference
(Eq.~\ref{eq:prob}), otherwise, agent $x$ will keep its strategy for the
next step.
\begin{equation}
    \label{eq:prob}
    p(s_x=s_y) = \frac{U_y-U_x}{8(T-P)},
\end{equation}
where $T$ is the temptation to defect and $P$ is the punishment for mutual
defection \cite{Cardinot2016ecta,Huang2015}.

\subsection{Experimental Set-up}
\label{sec:setup}

In this work, the population size is constant, $N=102 \times 102$, in all
simulations, which are run for $10^6$ Monte Carlo steps. In order to alleviate
the effect of randomness in the approach, each specific experimental set-up is
run 10 times.

Initially, we identify scenarios of intransitivity in the Optional Prisoner's
Dilemma game, i.e., the values of $b$, $l$ and $\Delta$ which promote
the coexistence of the three strategies (Sec.~\ref{sec:cyclic}). After that, the
stable population is tested to find the necessary conditions to break the
equilibrium. To do this, the following experiments are performed:
investigating the outcomes of populations with only two types of strategies (Sec.~\ref{sec:two});
exploring the effects of different mutation rates (Sec.~\ref{sec:three}); and
investigating the importance of the coevolutionary model in the sustenance of cyclic competition (Sec.~\ref{sec:destroy}).

\section{Cyclic Competition with Three Strategies}
\label{sec:cyclic}

Given an initial population with the same number of abstainers, cooperators and
defectors uniformly distributed, we start by investigating some parameter
settings in which a state of cyclic competition can be observed. Specifically,
we look for combinations of the loner's payoff $l$, the temptation to defect
$b$, and the link amplitude $\Delta/\delta$, which promote an equilibrium
between the three strategies. This experiment is based on the work described by
Huang et al. \cite{Huang2015}, where it is shown that some parameters settings
may promote cyclic dominance. However, it is noteworthy that this phenomenon is
not discussed in their paper and their methods consider only the classical
Prisoner's Dilemma.

In the traditional case of the Optional Prisoner's Dilemma game, i.e., for a
static and unweighted network ($\Delta = 0.0$), results show that abstainers or
defectors dominate in most scenarios and the dominance is closely related to
the payoff values.

In comparison, the population rarely reaches a state of cyclic competition
between the three strategies. In fact, it was only noticed when the temptation
to defect is in the range $b=[1.1,\ 1.2]$. Certainly, it is more likely to
happen in this scenario because there is no big advantage in choosing a
specific strategy and consequently the strategies tend to remain in
equilibrium. However, this behaviour is not very stable.

\begin{figure}[htb]
    \centering
    {\includegraphics[width=0.59\textwidth]{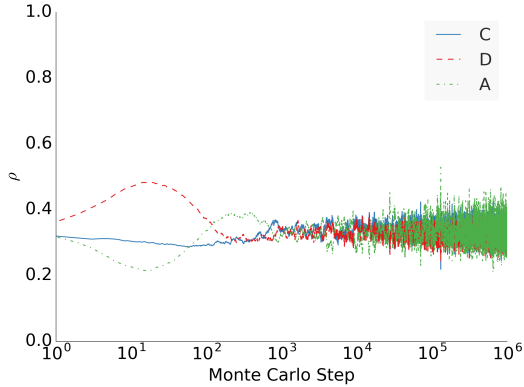}}
    \caption{
        Progress of the fraction of cooperators, defectors and abstainers
        during a Monte Carlo simulation for $\Delta/\delta=0.3$, $\delta=0.8$,
        $l=0.5$ and $b=1.9$.
    }
    \label{fig:mcs}
\end{figure}

When it comes to the cases of $\Delta > 0.0$, results show that a wide number
of different parameter settings can spontaneously promote the intransitive
behaviour, in which cooperators, defectors and abstainers remain in
equilibrium. For instance, Figure~\ref{fig:mcs} shows the progress, over the
Monte Carlo time steps, of each strategy for $\Delta/\delta=0.3$, $\delta=0.8$,
$l=0.5$ and $b=1.9$. We can observe that agents quickly organise in the
population in a way that the fraction of each strategy remains about
$33\% (\pm 7\%)$.

\begin{figure}[tb]
    \centering
     \begin{subfigure}[t]{0.4\textwidth}
        \includegraphics[width=\textwidth]{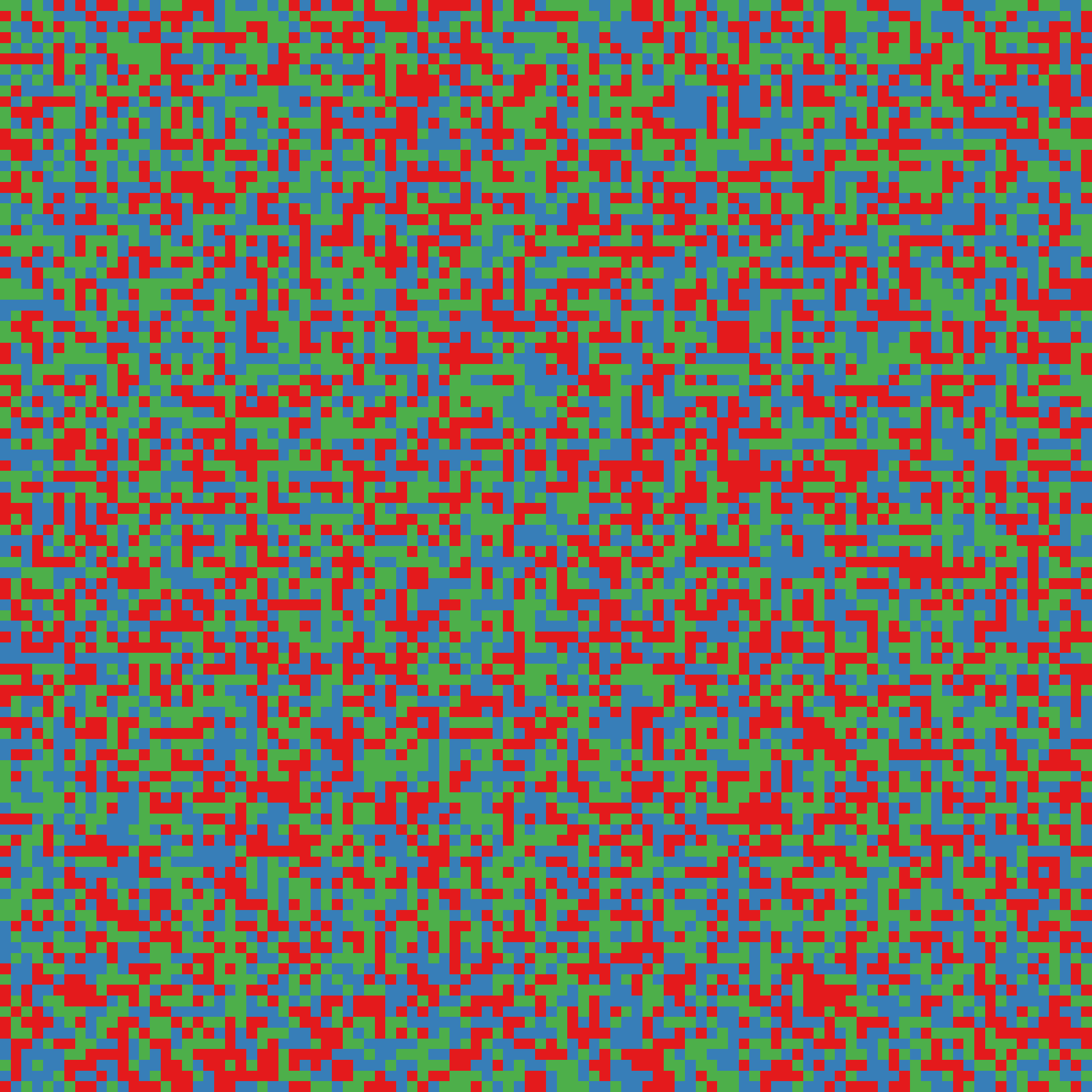}
        \caption{MC step $0$}
        \label{fig:snp_0}
    \end{subfigure}
    ~
     \begin{subfigure}[t]{0.4\textwidth}
        \includegraphics[width=\textwidth]{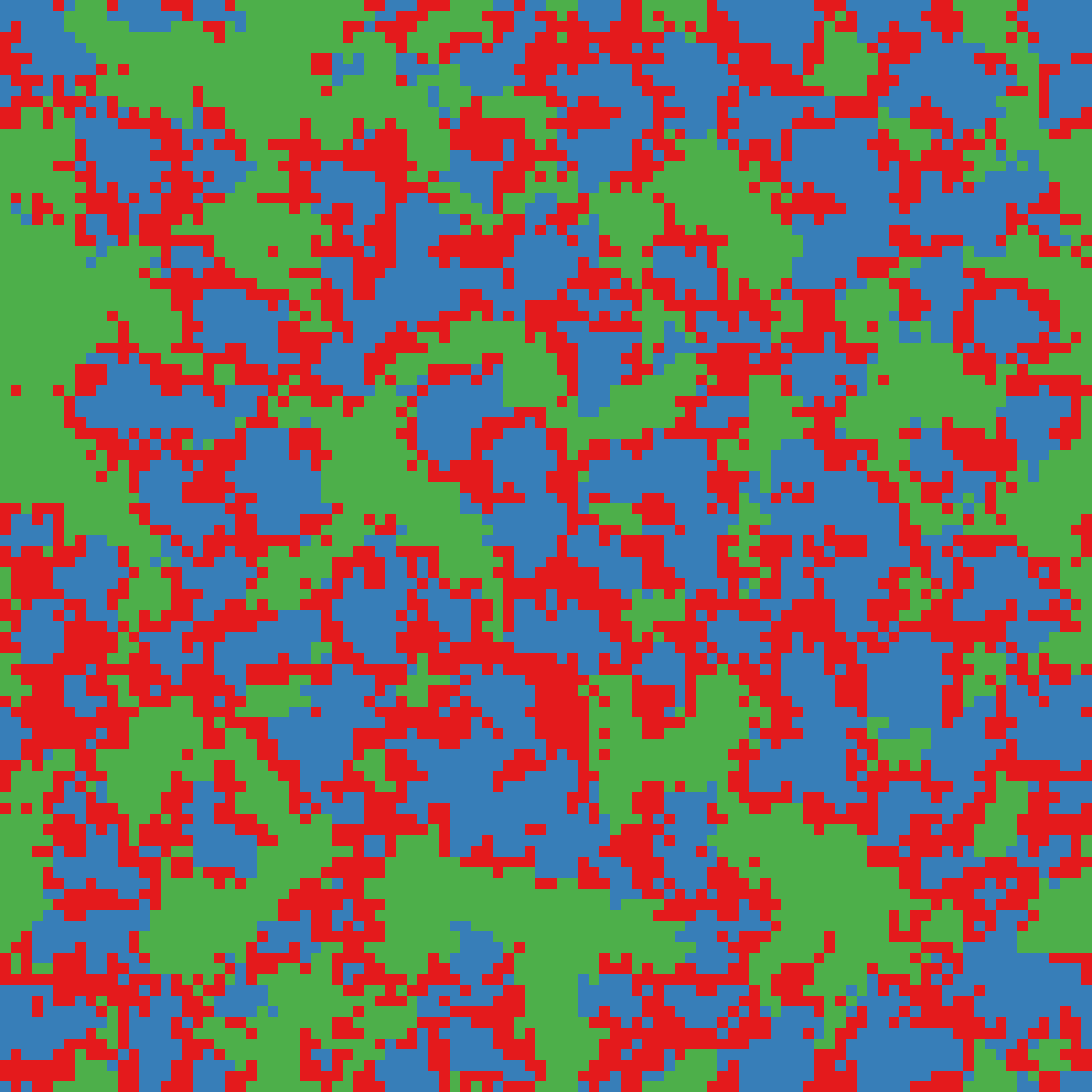}
        \caption{MC step $10^6$}
        \label{fig:snp_106}
    \end{subfigure}
    \caption{
        Snapshots of the distribution of the strategy in the Monte Carlo
        simulation for $\Delta/\delta=0.3$ $\delta=0.8$, $l=0.5$ and $b=1.9$.
        In this Figure, cooperators, defectors and abstainers are represented
        by the colours blue, red and green respectively.
    }
    \label{fig:snp}
\end{figure}

In order to investigate what spatial patterns emerge in this scenario, we also
took some snapshots of the population at different Monte Carlo steps.
Particularly, the MC steps $0$ and $10^6$ are illustrated in
Figure~\ref{fig:snp}.

An interesting phenomenon in this simulation is that abstainers tend to form
bigger clusters by dominating defectors. However, abstainers ensure that a small
fraction of defectors remain in its surrounding area as a mechanism to protect
them against an invasion of cooperators. Defectors, in turn, attempt to
encircle cooperators, disconnecting them from abstainers, as a way to isolate
and easily dominate them. These dynamics explain the reason why the population
is never fully dominated by any one strategy.

\section{Breaking the Cyclic Competition: Extinction of Species}
\label{sec:extinction}

In real-world systems, species may become extinct due to a variety of causes
such as climate change, habitat degradation, diseases, genetic factors, etc.
This sort of scenario is also present in many other domains, for example, in
business with the extinction of companies caused by a pricing war.

In other cases, species may mutate in order to avoid extinction. For example,
in the case of the common side-blotched lizards, it is known that
yellow-throated males can, in specific instances, mutate into blue-throated
males and once they transform, it cannot be reverted. Moreover, the
yellow-throated male is the only species able to undergo mutation
\cite{Sinervo2000}.

Inspired by the behaviour of some species in nature such as the side-blotched
lizards and the \textit{Escherichia Coli} bacteria, we investigate scenarios in
which a species can mutate and we investigate the effect that this phenomenon
can have on the population and its evolutionary mechanisms. These set of
experiments aim to explore what are the necessary conditions to maintain the
cyclic competition between the three strategies, even in adverse scenarios in
which the fraction of agents of a given strategy is reduced in the
population, up to its complete extinction.

\subsection{Two Species}
\label{sec:two}

For the experiments involving only two strategies, or species, we use the
evolved population (strategies and spatial structure) obtained in our first
experiment (Sec.~\ref{sec:cyclic}), i.e., the outcome of $10^6$ Monte Carlo
steps. All other parameter settings are kept the same ($\Delta/\delta=0.3$,
$\delta=0.8$, $l=0.5$ and $b=1.9$).

Before the simulation, we replace all individuals of a strategy by another,
obtaining a population of two strategies. It was found that a state of cyclic
competition between only two strategies cannot be reached and that the outcomes
are always a full dominance of a specific strategy. The results can be
summarised as follows:
\begin{itemize}
    \item With an initial population of \textit{C} and \textit{A}, \textit{C} will dominate.
    \item With an initial population of \textit{D} and \textit{A}, \textit{A} will dominate.
    \item With an initial population of \textit{C} and \textit{D}, \textit{D} will dominate.
\end{itemize}

As well as mimicking outcomes that we observe in nature, these results
highlight the importance of cyclic competition in the sustenance of
biodiversity. For instance, in our abstract model, the state of complete
dominance of a strategy does not necessarily incur advantages. In other words,
some species may prefer to live in smaller numbers in the environment in order
to give more opportunity for prey to develop. Thus, although defectors prefer
to stay away from abstainers, a population fully occupied by defectors lacks
resources ($P=0$). Thus, defectors need abstainers to keep the cooperators
alive, which in turn will enable defectors to increase their payoffs.

\subsection{Three Species}
\label{sec:three}

Following the same procedures performed for the pairwise simulation
(Sec.~\ref{sec:two}), and for the same parameter settings, we now explore the
behaviour of a population with three strategies, or species, after an adverse
scenario occurs in which most of the agents of a particular strategy undergo
mutation.

This analysis is important in order to understand the necessary conditions for
a population to maintain the stable coexistence of strategies in cyclic
competition. Thus, considering that the initial population has about $33\%$ of
each strategy, we run several Monte Carlo simulations varying the mutation rate
from $1$ to $99$ percent. Surprisingly, results show that the state of cyclic
competition is very robust and for all simulations the population quickly
returns to the equilibrium of about $33\%$ of each strategy with the same
spatial pattern as shown in Figure~\ref{fig:snp}.

\begin{figure}[tb]
    \centering
     \begin{subfigure}[t]{0.31\textwidth}
        \includegraphics[width=\textwidth]{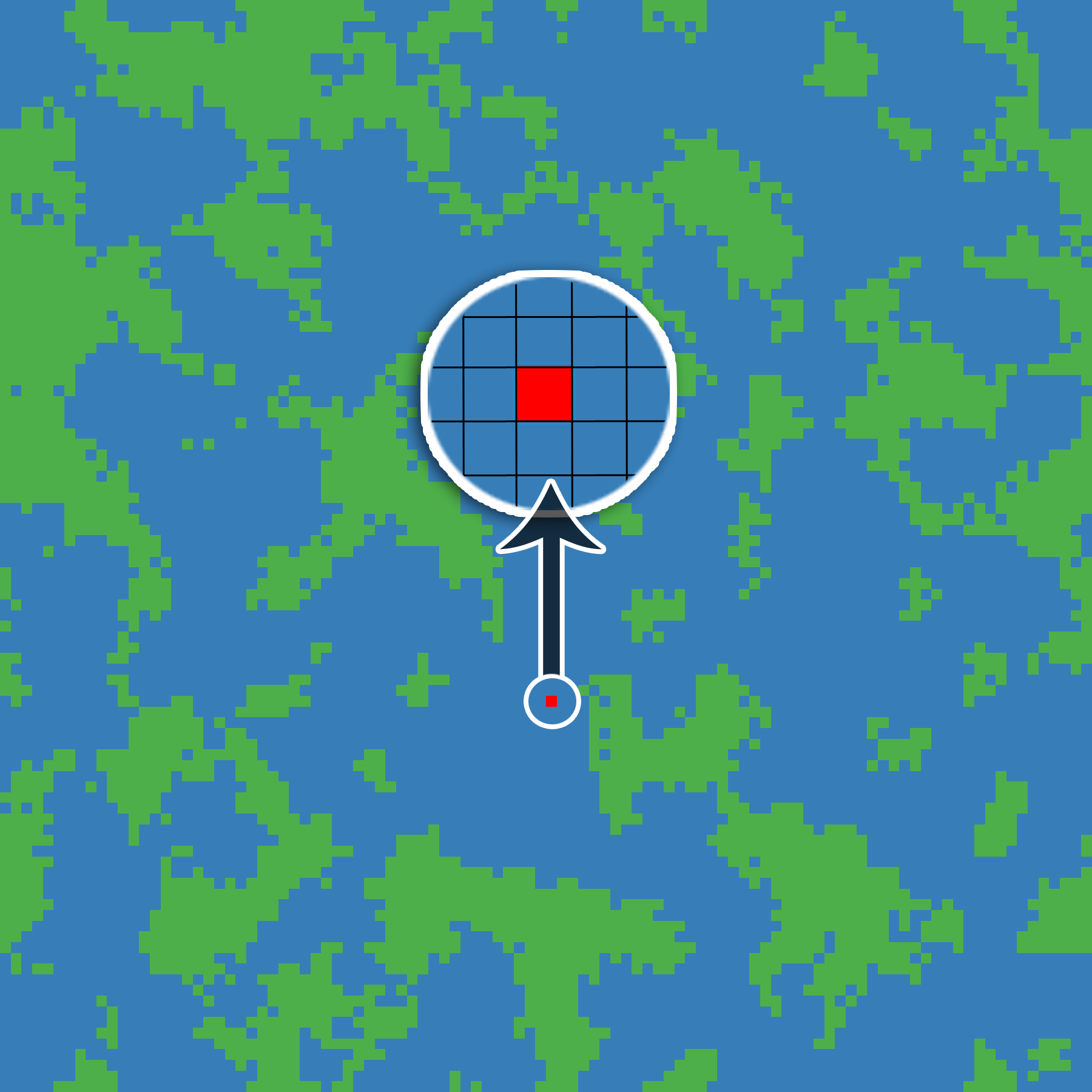}
        \caption{D into C}
        \label{fig:dc}
    \end{subfigure}
    ~
    \begin{subfigure}[t]{0.31\textwidth}
        \includegraphics[width=\textwidth]{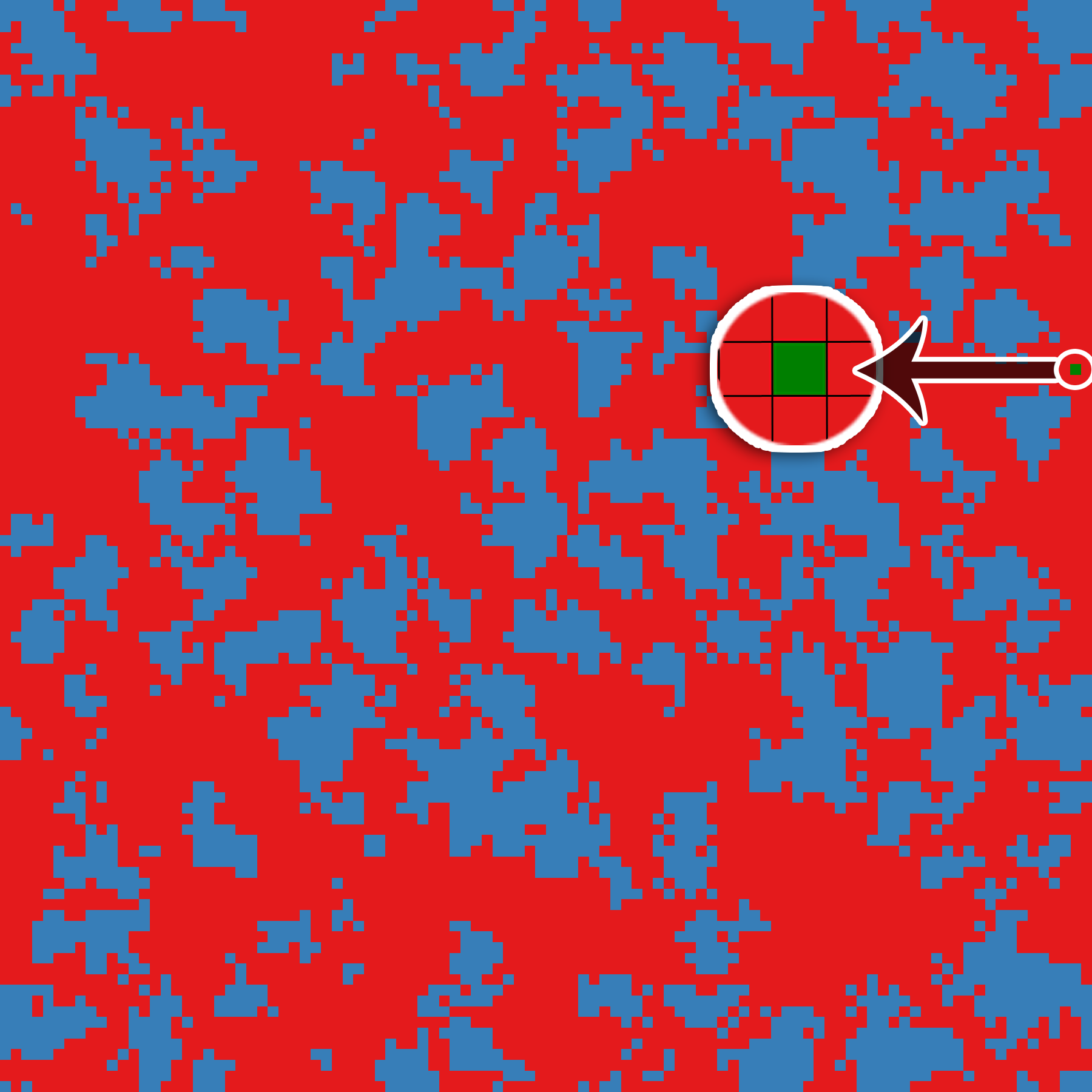}
        \caption{A into D}
        \label{fig:ad}
    \end{subfigure}
   ~
    \begin{subfigure}[t]{0.31\textwidth}
        \includegraphics[width=\textwidth]{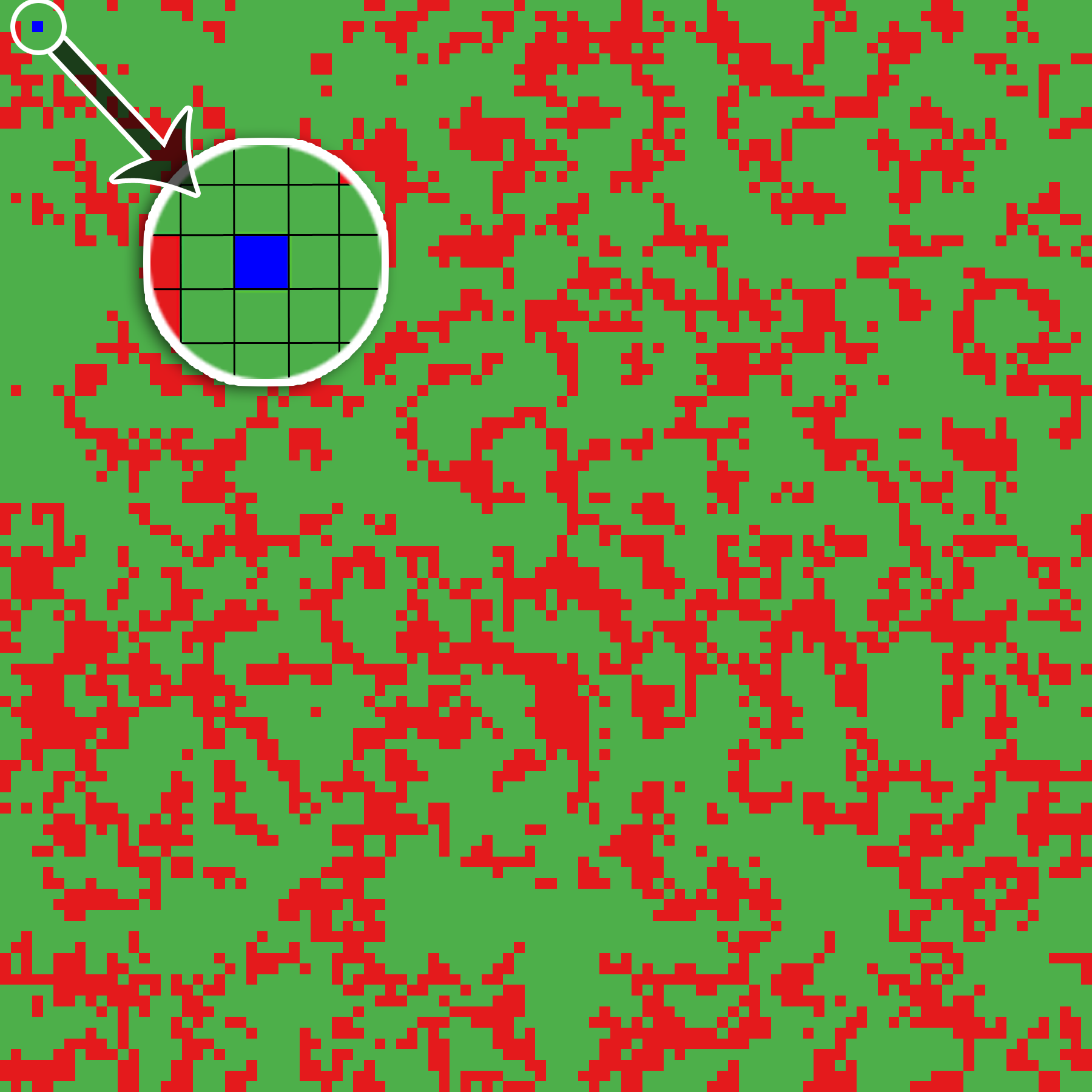}
        \caption{C into A}
        \label{fig:ca}
    \end{subfigure}
    ~
    \begin{subfigure}[t]{0.31\textwidth}
        \includegraphics[width=\textwidth]{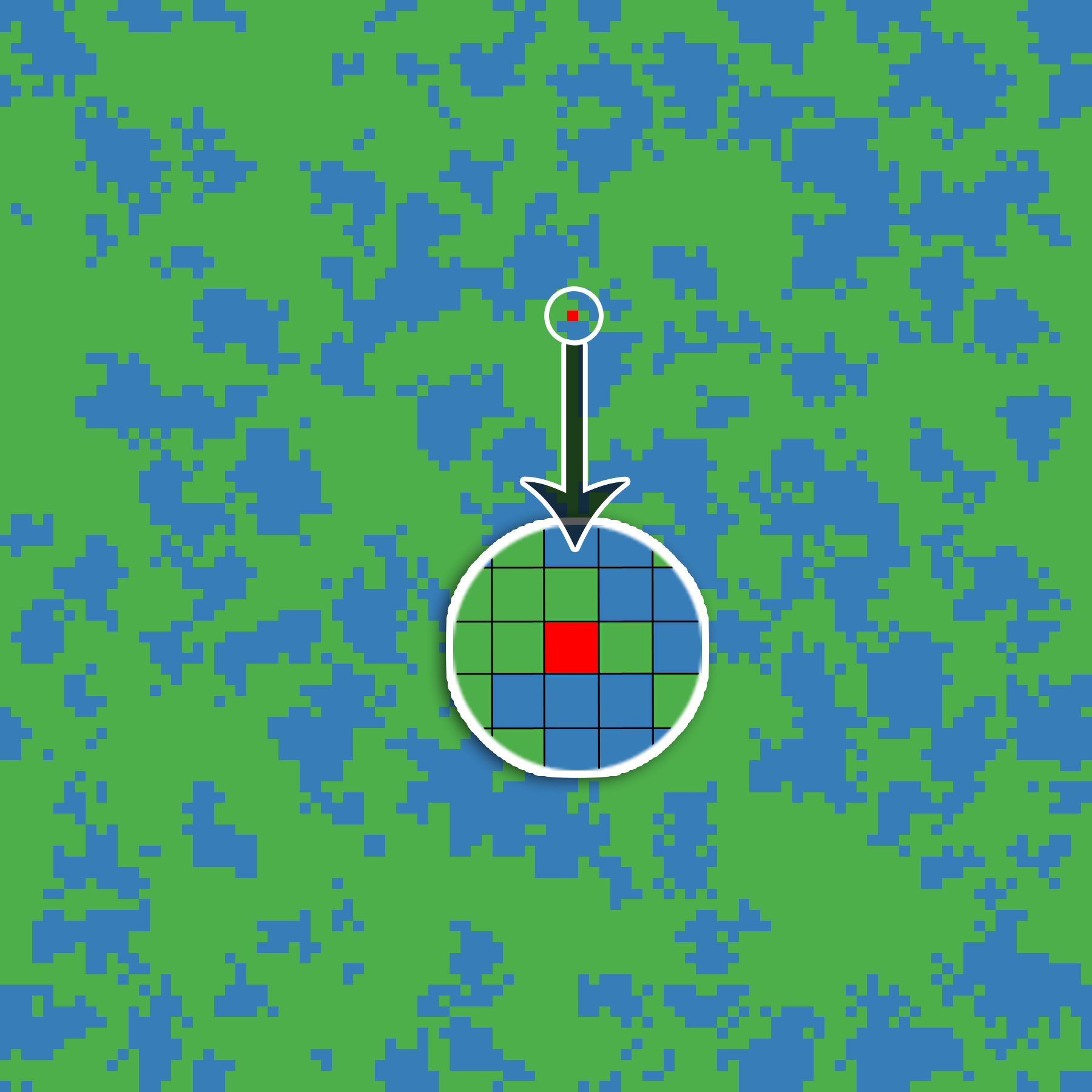}
        \caption{D into A}
        \label{fig:da}
    \end{subfigure}
    ~
    \begin{subfigure}[t]{0.31\textwidth}
        \includegraphics[width=\textwidth]{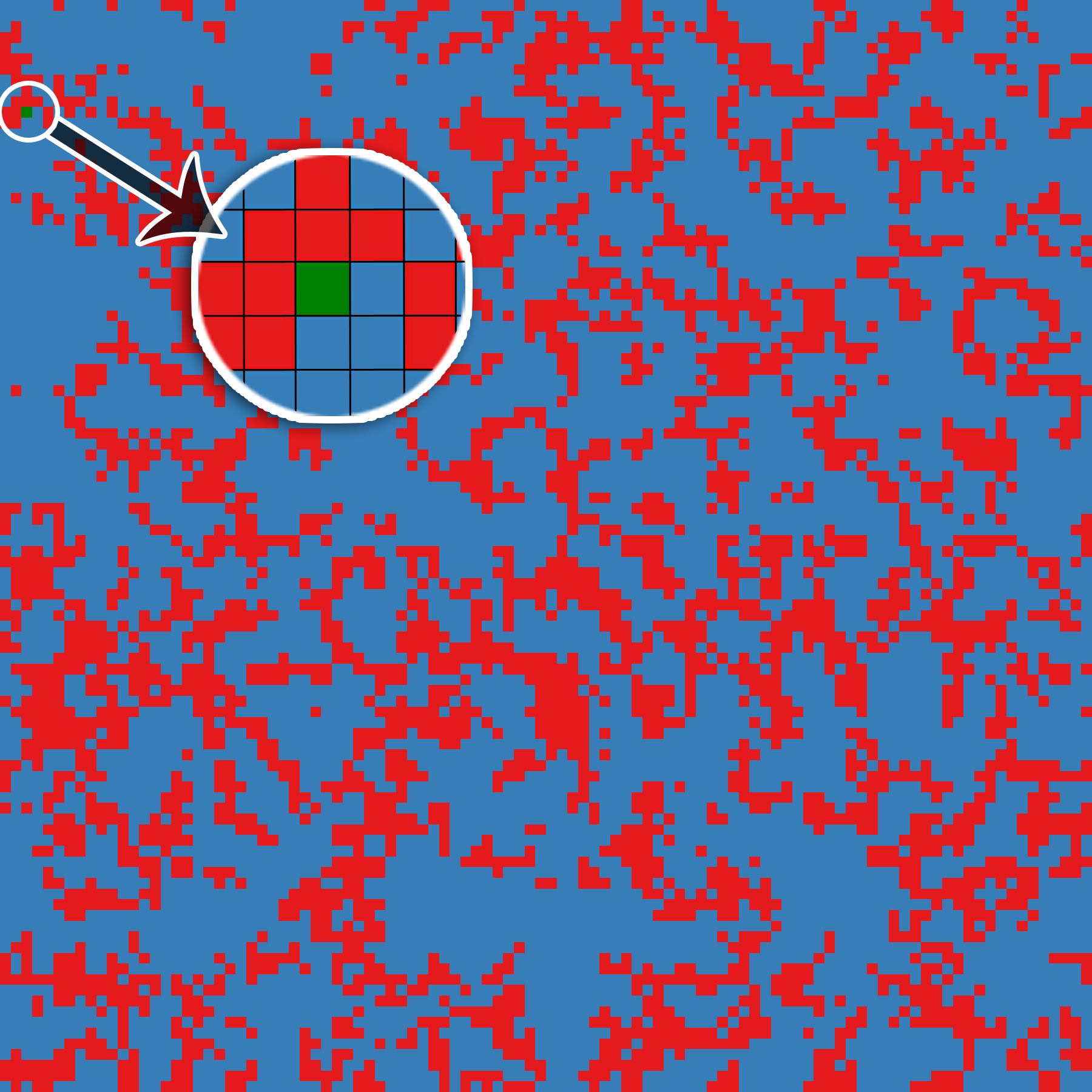}
        \caption{A into C}
        \label{fig:ac}
    \end{subfigure}
    ~
    \begin{subfigure}[t]{0.31\textwidth}
        \includegraphics[width=\textwidth]{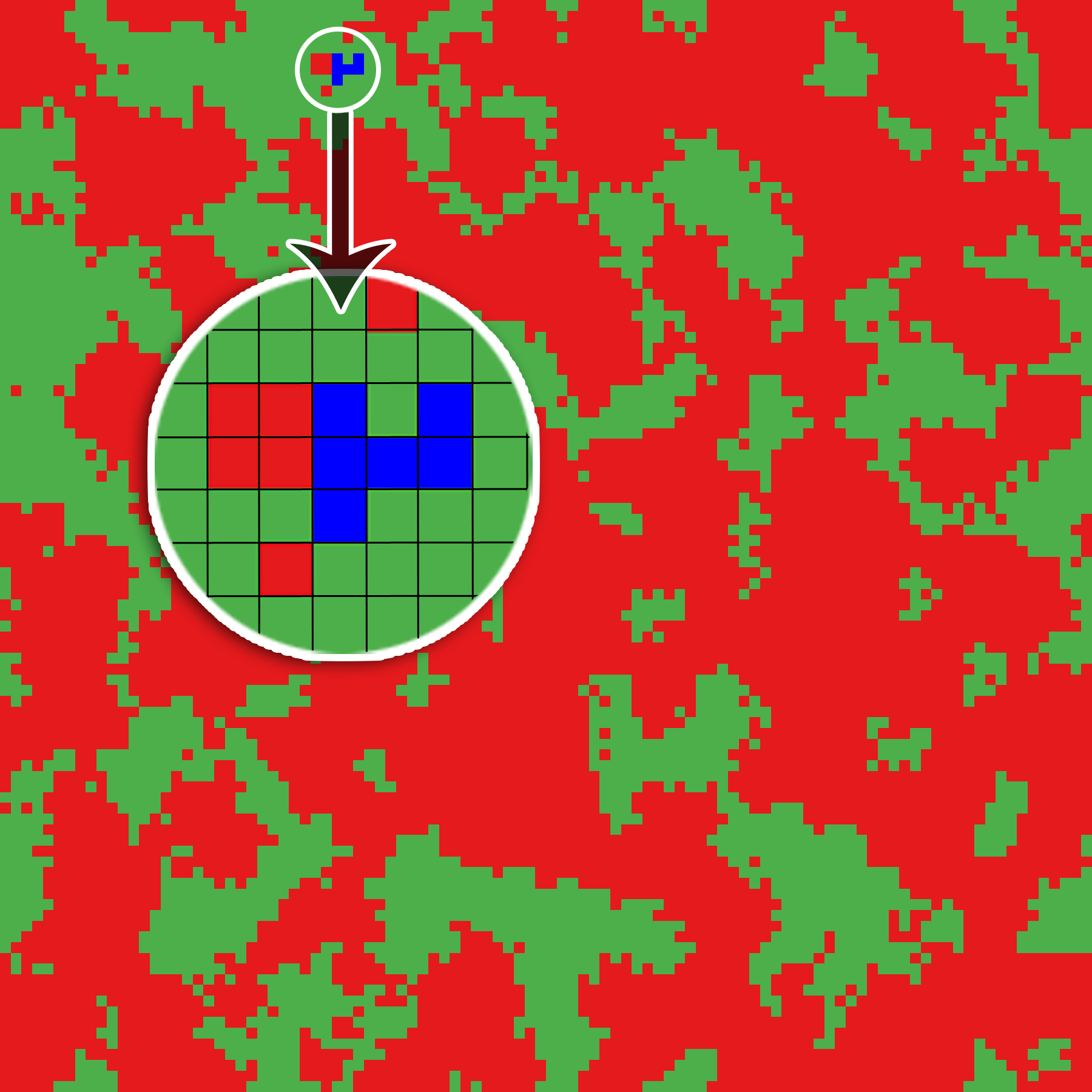}
        \caption{C into D}
        \label{fig:cd}
    \end{subfigure}
    \caption{
        Initial population after mutation of strategies. All scenarios recover
        the balance between the three strategies, showing a spatial pattern
        similar to the one illustrated in Figure~\ref{fig:snp}. In this Figure,
        cooperators, defectors and abstainers are represented by the colours
        blue, red and green respectively.
    }
    \label{fig:mutations}
\end{figure}

In order to further explain the results witnessed in these experiments, in
which the vast majority of our simulations with extremely high mutation rates,
i.e. $99.9\%$, converged to the equilibrium; we decided to analyse cases where
only one agent of a specific strategy does not mutate in the population (i.e.,
maximum mutation rate). This scenario, in turn, is illustrated in
Figure~\ref{fig:mutations}, which shows the snapshots of the mutated population
at the initial Monte Carlo step.

Results show that only one individual of a species needs to be kept in the
population in order for the cyclic competition to remain. All scenarios in
Figure~\ref{fig:mutations} quickly reverted back to the spatial pattern
illustrated in Figure~\ref{fig:snp}. However, it is only possible if the single
agent is linked with a sufficient number of subordinate agents. Namely, a
cooperator must have abstainers in its neighbourhood; similarly abstainers must
have defectors, and defectors must have cooperators. This condition explains
the robustness of our previous experiment, in which the balance was recovered
even when $99\%$ of the agents of a strategy were mutated. In this situation,
as we have a population of $10404$ agents, the chances of at least one of $104$
agents being connected to a satisfactory number of subordinated strategies is
very high.

It can be seen that when mutating $Ds$ into $Cs$, we unbalance the final
population to about $66.666\%$ $Cs$ and $33.333\%$ $As$. We observe that this
difference is not too relevant for the final outcome and that the key role in
sustaining the cyclic competition is in the hands of the single agent ($D$ in
this example). Results show that balance between the three strategies is always
recovered when the single agent is completely surrounded by its subordinate
strategy. For instance, a single defector surrounded by cooperators will never
be replaced by a cooperator because its utility will be greater than the
utility of any neighbour.

When the mutation occurs from a dominant to a subordinate strategy such as,
from defector to cooperator, abstainer to defector or cooperator to abstainer,
the chances of the single agent being kept in the middle of a suitable cluster
(subordinate strategy) are very high. This scenario can be observed in Figures
\ref{fig:dc}, \ref{fig:ad} and \ref{fig:ca}. However, in the opposite case it
is not generally possible because, during the evolutionary dynamics, a
subordinate strategy will almost never survive alone in the middle of a cluster
of a dominant strategy.

Therefore, the special aforementioned cases are not as stable as the case in
which the single agent is surrounded by subordinated neighbours. Its stability
will depend on the number of links with subordinated strategies and the
respective values of the link weights. In our experiments, it was observed that
when the single agent is a defector, the cyclic competition is recovered more
often when the number of cooperators in the neighbourhood is greater than or
equal to four (Fig.~\ref{fig:da}). When the single agent is an abstainer, at
least five of the eight neighbours need to be defectors (Fig.~\ref{fig:ac}).

Finally, as shown in Figure~\ref{fig:cd}, it is usually impossible to sustain
cyclic competition when the single agent kept after mutation is a cooperator.
It happens because cooperators are very sensitive to the presence of defectors
and, in this sort of scenario, they need at least one partner to be able to be
rewarded for mutual cooperation ($R$). Thus, better results are obtained as the
number of cooperators kept together increases.

\subsection{Destroying the Environment}
\label{sec:destroy}

Previous experiments discussed in Sections \ref{sec:two} and \ref{sec:three}
have considered the mutation of species (strategies) applied to an already
coevolved population (species and spatial environment). Namely, strategies
are transformed, but the spatial environment is still the same.

In this section we are interested in exploring the impacts of changing
the spatial environment at the mutation step such
that all link weights between agents are reset back to $w=1$ as existed in
the initial settings of the first experiment (Sec. \ref{sec:cyclic}).

Results show that the robustness witnessed in previous simulations, in which
only one species was needed to recover the balance between the three species,
is actually only possible because the spatial environment is kept unchanged.
The coevolutionary rules adopted enables agents to constantly strengthen
beneficial connections and weaken harmful ones, adapting the environment to fit
individual needs. Hence, in the previous experiments, as the non-mutated agent
is probably the most adapted in the neighbourhood, its strategy spreads
quickly, allowing the population to recover the balance.

This finding highlights the importance of the coevolutionary model in allowing
agents to adapt the environment to sustain the diversity of strategies.

\section{Conclusions and Future Work}
\label{sec:conclusion}

In this paper, we have investigated the phenomenon of cyclic dominance in a
coevolutionary Optional Prisoner's Dilemma, in which both game strategies and
edges linking agents adaptively evolve over time. An agent-based Monte Carlo
simulation approach was adopted to perform the evolutionary game in a
population of agents placed on a lattice grid with a Moore neighbourhood.

Despite the fact that the rock-paper-scissors game remains one of the most
oft-studied games in scenarios of cyclic dominance, in this paper, we showed
that the same behaviour can also be noticed in the Optional Prisoner's Dilemma
game, which allows the investigation of more complex scenarios that may lead to
a variety of outcomes.

Results show that cyclic dominance between the three strategies can emerge
spontaneously in a wide range of parameter settings, i.e., $l$, $b$, $\Delta$
and $\delta$, including the traditional case ($\Delta=0.0$) for an unweighted
and static network. However, it was observed that populations of only two
strategies can quickly lead to dominance of one strategy, which may lead to a
much lower performance.

Experiments revealed that the equilibrium between the three strategies is
maintained even in adverse scenarios, in which the mutation rate is extremely
high. It was observed that having only one agent of a strategy is often enough
to enable the population to revert back to a balanced state. However, this
single agent must be surrounded by subordinated agents, i.e., cooperators
surrounded by abstainers, abstainers surrounded by defectors and defectors
surrounded by cooperators.

Moreover, it was shown that the coevolutionary spatial method adopted plays a
key role in the sustenance of coexistence because it allows agents to also
adapt the environment, which is reasonable in more realistic scenarios. For
instance, in real life, a population is often changing the environment over time
in order to improve their performance and welfare. Thus, in adverse scenarios,
it is much easier for an individual to overcome and survive in such an
evolved environment.

Future work will involve the mathematical analysis of the necessary conditions
to sustain the coexistence of three competing strategies of the Optional
Prisoner's Dilemma game, allowing us to further explain the results obtained by
Monte Carlo simulations.

\vspace{3mm}
\noindent\textbf{Acknowledgement.}
This work was funded by CNPq-Brazil.

\bibliographystyle{splncs03}
\bibliography{aics2016}

\begin{thebibliography}{10}
\providecommand{\url}[1]{\texttt{#1}}
\providecommand{\urlprefix}{URL }

\bibitem{Cardinot2016sab}
Cardinot, M., Gibbons, M., O'Riordan, C., Griffith, J.: {Simulation of an
  Optional Strategy in the Prisoner's Dilemma in Spatial and Non-spatial
  Environments}, pp. 145--156. Springer International Publishing, Cham (2016)

\bibitem{Cardinot2016ecta}
Cardinot, M., O'Riordan, C., Griffith, J.: {The Optional Prisoner's Dilemma in
  a Spatial Environment: Co-evolving Game Strategy and Link Weights}. In:
  Proceedings of the 8th IJCCI (In Press) (2016)

\bibitem{Cheng2014}
Cheng, H., Yao, N., Huang, Z.G., Park, J., Do, Y., Lai, Y.C.: Mesoscopic
  interactions and species coexistence in evolutionary game dynamics of cyclic
  competitions. Sci Rep  4 (2014)

\bibitem{Demirel2011}
Demirel, G., Prizak, R., Reddy, P.N., Gross, T.: Cyclic dominance in adaptive
  networks. EPJ B  84(4),  541--548 (2011)

\bibitem{Hauert2008}
Hauert, C., Traulsen, A., Brandt, H., Nowak, M.A.: {Public goods with
  punishment and abstaining in finite and infinite populations}. Biol Theory
  3(2),  114--122 (2008)

\bibitem{Huang2015}
Huang, K., Zheng, X., Li, Z., Yang, Y.: {Understanding Cooperative Behavior
  Based on the Coevolution of Game Strategy and Link Weight}. Sci. Rep.  5,
  14783 (2015)

\bibitem{Jeong2014}
Jeong, H.C., Oh, S.Y., Allen, B., Nowak, M.A.: Optional games on cycles and
  complete graphs. J Theor Biol  356,  98--112 (2014)

\bibitem{Jorgensen2014}
Jorgensen, S., Fath, B.: {Encyclopedia of Ecology}. Elsevier (2014)

\bibitem{Juul2013}
Juul, J., Sneppen, K., Mathiesen, J.: Labyrinthine clustering in a spatial
  rock-paper-scissors ecosystem. Phys Rev E  87,  042702 (2013)

\bibitem{Kerr2002}
Kerr, B., Riley, M.A., Feldman, M.W., Bohannan, B.J.: Local dispersal promotes
  biodiversity in a real-life game of rock--paper--scissors. Nature  418(6894),
   171--174 (2002)

\bibitem{Knebel2015}
Knebel, J., Weber, M.F., Kr{\"u}ger, T., Frey, E.: Evolutionary games of
  condensates in coupled birth-death processes. Nature  6 (2015)

\bibitem{Mobilia2010}
Mobilia, M.: {Oscillatory dynamics in rock–paper–scissors games with
  mutations}. J Theor Biol  264(1),  1--10 (2010)

\bibitem{Nowak1992}
Nowak, M.A., May, R.M.: Evolutionary games and spatial chaos. Nature
  359(6398),  826--829 (1992)

\bibitem{Perc2010}
Perc, M., Szolnoki, A.: {Coevolutionary games -- A mini review}. Biosystems
  99(2),  109--125 (2010)

\bibitem{Reichenbach2007}
Reichenbach, T., Mobilia, M., Frey, E.: Mobility promotes and jeopardizes
  biodiversity in rock--paper--scissors games. Nature  448(7157),  1046--1049
  (2007)

\bibitem{Sinervo2000}
Sinervo, B., Miles, D.B., Frankino, W., Klukowski, M., DeNardo, D.F.:
  {Testosterone, Endurance, and Darwinian Fitness: Natural and Sexual Selection
  on the Physiological Bases of Alternative Male Behaviors in Side-Blotched
  Lizards}. Horm Behav  38(4),  222--233 (2000)

\bibitem{Smith1982}
Smith, J.M.: Evolution and the theory of games. CUP, Cambridge (1982)

\bibitem{Szabo2002}
Szab\'o, G., Hauert, C.: {Evolutionary Prisoner's Dilemma games with voluntary
  participation}. Phys Rev E  66,  062903 (2002)

\bibitem{Szolnoki2014}
Szolnoki, A., Mobilia, M., Jiang, L.L., Szczesny, B., Rucklidge, A.M., Perc,
  M.: Cyclic dominance in evolutionary games: a review. J R Soc Interface
  11(100) (2014)

\bibitem{Szolnoki2010}
Szolnoki, A., Wang, Z., Wang, J., Zhu, X.: {Dynamically generated cyclic
  dominance in spatial Prisoner’s Dilemma games}. Phys Rev E  82(3),  036110
  (2010)

\bibitem{Wang2014}
Wang, Z., Szolnoki, A., Perc, M.: {Self-organization towards optimally
  interdependent networks by means of coevolution}. New J Phys  16(3),  033041
  (2014)

\bibitem{Yu2015}
Yu, Q., Chen, R., Wen, X.: {Evolutionary Voluntary Prisoner’s Dilemma Game
  under Deterministic and Stochastic Dynamics}. Entropy  17(4),  1660 (2015)

\bibitem{Yu2016}
Yu, Q., Fang, D., Zhang, X., Jin, C., Ren, Q.: {Stochastic Evolution Dynamic of
  the Rock--Scissors--Paper Game Based on a Quasi Birth and Death Process}. Sci
  Rep  6 (2016)

\bibitem{Zhou2016}
Zhou, H.J.: The rock–paper–scissors game. Contemp Phys  57(2),  151--163
  (2016)

\bibitem{Zimmermann2001}
Zimmermann, M.G., Egu{\'i}luz, V.M., Miguel, M.S.: {Cooperation, Adaptation and
  the Emergence of Leadership}, pp. 73--86. Springer, Berlin, Heidelberg (2001)

\end{thebibliography}

\end{document}